\documentclass{elsart3-1}

\usepackage{amssymb}
\usepackage{bm}% bold math
\usepackage{amsmath,amssymb}
\usepackage[english,francais]{babel}

\newtheorem{e-proposition}[theorem]{Proposition}

\newtheorem{e-definition}[theorem]{Definition\rm}

\renewcommand{\theequation}{\arabic{equation}}
\def\og{\leavevmode\raise.3ex\hbox{$\scriptscriptstyle\langle\!\langle$~}}
\def\fg{\leavevmode\raise.3ex\hbox{~$\!\scriptscriptstyle\,\rangle\!\rangle$}}

\renewcommand{\v}[1]{\underline{#1}{}}
\newcommand{\vv}[1]{\bm{#1}}

\newcommand{\tun}{\vv{\delta}}

\newcommand{\cauchy}{\vv{\sigma}}
\newcommand{\epsl}{{\vv{\varepsilon}}}
\newcommand{\vvvv}[1]{\mathbb{#1}{}}

\newcommand{\Elas}{\vvvv{C}}
\newcommand{\cauchytilde}{\vv{\tilde \sigma}}
\newcommand{\vn}{\v{n}}

\begin{document}

\begin{frontmatter}


% \author{Name\thanksref{label2}}
\selectlanguage{english}
\title{Elastic modulus of a colloidal suspension of rigid spheres at
  rest}

% The [label2] can be deleted if all authors share the same address

\selectlanguage{english}
\author[authorlabel1]{Laurentiu Pasol},
\ead{laurentiu.pasol@espci.fr}
\author[authorlabel2]{Xavier Chateau}
\ead{xavier.chateau@lcpc.fr}

\address[authorlabel1]{Laboratoire Physique Thermique-ESPCI, 10 rue
  Vauquelin 75231 Paris cedex 05, France.}
\address[authorlabel2]{Institut Navier, Laboratoire des Mat\'eriaux et
  des Structures du G\'enie Civil,  
2 all\'ee Kepler 77420 Champs sur Marne, France.}

% if you know the dates of reception, and acceptation you can put them now;
% idem for the name of the person presenting the Note

\medskip
\begin{center}
{\small Received 20 june 2007; accepted after revision 19 February 2008\\
Presented by Andr\'e Zaoui}
\end{center}

\begin{abstract}
% Text of abstract in English
By modeling a colloidal suspension at rest as a solid, a new expression for 
the linear elastic modulus is obtained. 
This estimate is valid for a yield stress colloidal suspension
submitted to a small strain.
Interestingly, it is also possible to perform hypothesis allowing to
recover the high-frequency  modulus classically found by means of a
classical 'fluid approach'. 
However, in most of the situations, the moduli obtained by the two
approaches are different.
{\it To cite this article: L. Pasol, X. Chateau, C. R.
Mecanique 336 (2008) 512-517.}

\vskip 0.5\baselineskip

\selectlanguage{francais}
% Text of abstract in French
\noindent{\bf R\'esum\'e}
\vskip 0.5\baselineskip
\noindent
{\bf Module d'\'elasticit\'e d'une suspension collo\"{i}dale de
  sph\`eres dures au repos.}
En mod\'elisant une suspension collo\"{i}dale au repos comme un
solide, on obtient une nouvelle expression pour le module
d'\'elasticit\'e lin\'eaire. 
Cette expression permet d'estimer le module d'une suspension
collo\"idale poss\'edant un seuil d'\'ecoulement soumise \`a une d\'eformation
infinit\'esimale.
On montre \'egalement que sous certaines hypoth\`eses, cette approche
permet de retrouver l'expression du module \'elastique \`a grande
fr\'equence obtenu par une approche classique de type fluide.
{\it Pour citer cet article~: L. Pasol, X. Chateau, C. R.
Mecanique 336 (2008) 512-517.}

%Now keywords/mots-clÈs, the first coming from the Cras Mecanique list
\vskip 0.5\baselineskip
\noindent{\small{\it Mots-cl\'es~:} rh\'eologie; suspension collo\"{i}dale; 
module d'\'elasticit\'e}}
\end{abstract}
\end{frontmatter}

% now the Version franÁaise abrÈgÈe, if it exists
%\selectlanguage{francais}
%\section*{Version fran\c{c}aise abr\'eg\'ee}
% Text of your Version franÁaise abrÈgÈe here

\selectlanguage{english}
% main text
\section{\label{intro}Introduction}

The macroscopic rheological properties of a colloidal suspension are the
counterpart at the macroscopic scale of phenomena occurring at the
length scale of the particles.
The forces applied to the particles originate from different
phenomena: colloidal forces, hydrodynamic interactions, Brownian
motions, etc.
The intensity of these forces depends upon several parameters as the
temperature, the size of the particles, the separation gap between
particles, or the pH of the suspending fluid.

At the macroscopic scale, the elastic modulus is one of the parameters
which characterize the suspension behavior. 
It has been recognized from a long time that its value depends in
particular upon the interactions between particles. 
The problem of
the transition from the microscale to the macroscale in
view of the prediction of the macroscopic properties of the suspension
has been the matter of intensive research for decades.
The main underlying motivation for this work is to predict the  overall
behavior of the suspension from the description of the particles properties. 
Most of the results have been obtained in the framework of
statistical physics.
For example Zwanzig and Mountain computed the high frequency shear modulus
of a simple monoatomic fluid where only binary interactions are likely
to occur in the absence of Brownian motion and hydrodynamic
forces~\cite{Zwanzig-Mountain:1965}.
Wagner generalized this results in order to account for Brownian
motion and hydrodynamic interaction by means of a linear
theory~\cite{Wagner:1993}.
He showed that Zwanzig and Mountain results are still valid when
hydrodynamic interactions between particles are negligible.
For concentrated suspensions Lionberger and Russel showed
that long-range hydrodynamic interactions can be neglected, the main
contribution coming from the lubrification
forces~\cite{Lionberger-Russel:1994}.
Brady developed in~\cite{Brady:1993} a theory accounting for specific
interparticle force laws and their influence on the suspension
rheology in the linear regime in the framework of a mechanical approach. 
Divergence of the viscosity at random close packing density was
recovered and the value of the exponent is predicted as a function of
the interparticles forces (Brownian hard spheres or particle
interacting through strongly repulsive colloidal forces).

The difficulty encounted to experimentally measure colloidal forces
also motivated interest in this field, mainly for elasticity.
The idea is to use theoretical results obtained in the framework of
change of scale methods in order to determine microscopic
properties of the suspension from the measurement of macroscopic
elastic modulus.
Such a method was proposed
in~\cite{Bergenholtz-Willenbacher-Wagner-Morrison-Ende-Mellema:1998}
to estimate the effective surface charge of particles of a
concentrated suspension from the measurement of the high-frequency
elastic modulus.

In most of these works, the suspension is modeled as a fluid.
From a practical point a view, it is well known that concentrated colloidal
suspensions often exhibit a yield stress: they flow and behave like a
fluid only when submitted to a stress above the yield stress.
Otherwise, the behave like a solid.
In the solid state, a colloidal suspension can be seen as a
disordered solid in which the particles form a connected network.
As long as the applied forces are not big enough to trigger off finite
displacement of the particles from their rest position, the
overall mechanical properties of the suspension can be predicted in
the framework of change of scale methods pertaining to heterogeneous
solid materials.

In this paper, we show that it is possible to perform such approach
in order to estimate elastic modulus for a colloidal suspension.
For simplicity, we restrict ourselves to the situations where
hydrodynamic interactions and Brownian motions are neglected.
As a consequence, our results are valid only for naught-velocity
loading in the solid regime (quasistatic elastic behavior) or
hight-frequency loading in the fluid regime.

The paper is organized as follows. 
We begin by recalling the relation linking the Cauchy stress tensor to
the interaction force.
Then, we compute the macroscopic strain-stress
behavior law in the linear regime.
Finally, a new estimate for the elastic shear modulus of
the suspension is obtained before we conclude.

\section{ Cauchy tensor for a colloidal suspension at rest}
\label{section2}
We consider a monodiperse suspension of spherical particles with
radius $a$ distributed in an incompressible Newtonian fluid. 
Particles interact through colloidal interaction forces.
The suspension is at rest both at the microscopic and the
macroscopic scales, so that hydrodynamic interaction forces are
negligible.
It is assumed that no forces from outside are applied on the particles
or on the fluid.

Consider a representative elementary volume (r.e.v.) of the suspension
occupying the geometrical domain $V$ and containing $N$ particles ($N
\gg 1$).
The macroscopic Cauchy stress tensor $\cauchy$ can be computed from the
knowledge of the microscopic quantities using the classical Batchelor
equation~\cite{Batchelor:1970,Batchelor:1977}:
\begin{equation}
  \label{eq:defContrainte}
  \cauchy  =  \frac{1}{V} \int_{V} \cauchytilde dV 
           =  -\frac{1}{V} \int_{V_{\text{f}}} p \tun dV +
\frac{1}{V} \sum_{i=1}^{N} \vv{s}_i
\end{equation}
where~$p$ denotes the fluid pressure, $\tun$ the second order unit
tensor,  $V_{\text{f}}$ the geometrical
domain filled by the fluid and $\vv{s}_i$ the stresslet of the
particle~$i$, equal to 
\begin{equation}
  \label{eq:defStresslet}
   \vv{s}_i   =  {\displaystyle \frac{1}{2} \int_{A_{i}} 
\left( \cauchytilde \cdot \vn
  \otimes \left(\v{x}- \v{x}_{i} \right) + \left(\v{x}- \v{x}_{i}
     \right)  \otimes \cauchytilde \cdot \vn \right)  dS}
    =  {\displaystyle \int_{A_{i}} \left(\v{x}- \v{x}_{i} \right)
    \otimes \cauchytilde \cdot \vn dS }
\end{equation}
with $\cauchytilde$, the cauchy stress tensor field in the particles,
$\v{x}_i$ the center of particle~$i$ and
$A_{i}$ the boundary of the domain occupied by the particle~$i$.
In equation~\ref{eq:defStresslet}, $\vn$ denotes the outer unit normal
to the domain $A_{i}$ and $\v{x}$ the position vector in the studied
configuration.

We consider only stable colloidal dispersions where interaction forces
dominate Brownian effects.
Then the particles form a disordered network.
The only contribution of particle~$i$ to the stress tensor is the
stresslet $\vv{s}_i$.
Thanks to the assumption that no external forces apply on the
particles, the strain tensor~$\cauchytilde$ is symmetric and so is the
macroscopic Cauchy stress tensor.
Furthermore, the fluid pressure is uniform over the domain occupied by
the fluid in the representative elementary volume.
As it is classical, it is assumed that interparticle forces derive
from a potential~\cite{Russel-Saville-Schowalter:1995}.
Then, $\v{F}_{j \rightarrow i}$, the force applied by particle~$j$ to
particle~$i$ reads $ \v{F}_{j \rightarrow i} = -\frac{\partial \psi}{\partial
  \v{x}_i}\left(\v{x}_{ij}\right)$ 
%\begin{equation}
%  \label{eq:Fjsuri}
%  \v{F}_{j \rightarrow i} = -\frac{\partial \psi}{\partial
%  \v{x}_i}\left(\v{x}_{ij}\right)
%\end{equation}
where  $\v{x}_{ij} = \v{x}_{j}-\v{x}_{i}$ denotes the vector
connecting the center of particle~$i$ to the center of particle~$j$. 
It is advisable to note here that when the interparticle forces are
described by force vectors, the Batchelor's
equation~\ref{eq:defContrainte} is no more valid.
Putting  the equilibrium equation for both the whole r.e.v. and each
particle in equation~\ref{eq:defContrainte} allows to compute the
macroscopic Cauchy stress tensor as a function of the interparticle
forces and the fluid
pressure~\cite{Batchelor:1977,Russel-Saville-Schowalter:1995,Chateau02}
\begin{equation}
  \label{eq:defContrainte2}
  \cauchy = -p \tun - \frac{1}{V} \sum_{i < j}^{N} \v{F}_{i \rightarrow
    j} \otimes \v{x}_{ij}
\end{equation}
This relation can also be obtained in the framework of a micromechanical
approach to the behavior of a heterogeneous material~\cite{Zaoui02}.
It is recalled that mechanical homogenization techniques aim at finding
the overall behavior of a system in a form of relationship between
macroscopic stress and strain tensors from the response of the
r.e.v. to a mechanical loading in which one of the two macroscopic
tensors acts like a loading parameter.
As the mechanical behavior of the suspension does not depend on the
value of the fluid pressure, it is assumed that $p$ is naught in the
sequel.

Of course, one has to consider an arbitrary realization of the
material system to compute the equation~\ref{eq:defContrainte2}.
In order to obtain results which do not depend on the particular
selected realization, it is necessary to average the
equation~\ref{eq:defContrainte2} over all the possible realizations
of the system.
Let $\mathcal{C}_{N}=\{\v{x}_{1},\v{x}_{2},...,\v{x}_{N} \}$ denotes a
particular realization for the centers of the $N$ particles embedded
in the r.e.v. and let $P_{N}(\v{x}_{1}, \v{x}_{2},...,\v{x}_{N})$
denotes the probability of finding simultaneously the particle centers 
in $\v{x}_{1},\v{x}_{2},...,\v{x}_{N}$.
As the particles are indistinguishable the probability to find
simultaneously the center of one particle in $\v{x}_{1}$ and the center
of another particle in $\v{x}_{2}$ reads
\begin{equation}
p_{2}(\v{x}_{1},\v{x}_{2})  =
\frac{1}{(N-2)!}\int_{V^{N-2}}P_{N}(\mathcal{C}_{N})d\v{x}_{3}\dots
d\v{x}_{N} 
= n^{2}g (\v{r})
\label{conddeux}
\end{equation}
where $\v{r}=\v{x}_{1}-\v{x}_{2}$ and $n=N/V$ denotes the number
density of particles in the representative elementary volume.
The second equality of equation~\ref{conddeux} is only valid for
statistically homogeneous  suspensions.
It is assumed that this condition is fulfilled in the sequel.
$g$ is the the radial distribution
function~\cite{Hansen-McDonald:1986}. 
Averaging the stress tensor equation~\ref{eq:defContrainte2} with the
probability $p_{2}(\v{x}_{1},\v{x}_{2})$ defined by~\ref{conddeux}
yields 
\begin{equation}
< \cauchy >= - \frac{n^{2}}{2}\int_{V}\v{r}\otimes \v{F} (\v{r})\,g(\v{r})\,
dV(\v{r})
\label{stressmacro}
\end{equation}  
where the convention $\v{F} (\v{r}) = \v{F}_{2\rightarrow 1}(\v{r})$
has be used to simplify the notations.

\section{Elastic modulus}
\label{section3}

In order to identify the tangent moduli of the suspension, a
macroscopic linearized strain $\epsl$ is applied to the representative
elementary
volume occupying the geometrical domain~$V_0$ in the undeformed
configuration.
A particle located at $\v{X}_i$ in the reference configuration
moves to the position $\v{x}_i$ in  the deformed configuration.
Such a macroscopic loading can be defined by the so-called Hashin
boundary condition according to which the displacement of particles
located on the boundary of the r.e.v. is prescribed, equal to $\epsl
\cdot \v{X}_i$.
As the particles are rigid, the macroscopic loading must comply with
the incompressibility condition $\tun : \epsl = 0$.
It is worth noting that for each realization of the suspension, local
material heterogeneities are responsible for microscopic fluctuations
of the displacement around the linear field $\epsl \cdot \v{X}$.

Up to the first order in~$\epsl$, the Cauchy stress tensor on the
deformed configuration reads
\begin{equation}
\label{eq:contraintesolide}
<\cauchy>  =  <\vv{\pi}^0>  -   \frac{n^2}{2} \int_{V_0} \left(\v{R} \otimes
            \v{F}  \left(\frac{d g_0}{d\v{R}} \cdot \mathbb{A} :
            \epsl\right) + \left(\mathbb{A}
            : \epsl \otimes \v{F}   + \v{R} \otimes
            \frac{d  \v{F}}{d\v{R}}  \cdot \mathbb{A}
            : \epsl\right) g_0 \right)\,dV
\end{equation}
with 
\begin{equation}
  \label{eq:defPio}
  <\vv{\pi}^0> =  -   {\displaystyle \frac{n^2}{2} \int_{V_0}  \v{R} \otimes
            \v{F}(\v{R}) \,  g_0(\v{R})\, dV} 
\end{equation}
$\vv{\pi}^0$ denotes the Piola-Kirchhoff stress tensor on the
underformed configuration~\cite{Germain73,Salencon2001}.
$<\vv{\pi}^0>$ is equal to the Cauchy stress tensor
in the undeformed configuration.
It is worth noting that quantities $\v{F}$, $g_0$ and $\mathbb{A}$
are function of the position vector $\v{R}$ (the dependence have been
omitted for simplicity).
The relative displacement concentration tensor defined by
$\mathbb{A}(\v{R}) = d \v{R} / d \epsl$ allows to compute the relative
displacement $\v{r} - \v{R} = \v{x}_2-\v{x}_1 -(\v{X}_2 - \v{X}_1)$ of
two particles induced by the loading~$\epsl$.

Explicitly knowing the third order tensor $\mathbb{A} (\v{R})$ would
allow to compute the behavior law linking the Cauchy stress tensor to
the linearized strain tensor~$\epsl$.
It is assumed in the sequel that the behavior of the suspension is
linear elastic at the macroscopic scale.
Then, the  macroscopic state
law reads~\cite{Salencon2001}
\begin{equation}
  \label{eq:cauchyelasdefopure}
  \cauchy = \vv{\pi}^0 
+ \epsl \cdot \vv{\pi}^0
+ \vv{\pi}^0 \cdot \epsl + \Elas : \epsl 
=  \vv{\pi}^0 + \vvvv{L} (\vv{\pi}^0) : \epsl
\end{equation}
where $\vvvv{L} (\vv{\pi}^0)$ denotes  the tangent tensor and $\Elas$
the elastic tensor.
It is recalled that the tensor  $\vvvv{L} (\vv{\pi}^0)$ is generally not
equal to the elastic tensor $\Elas$ and does not satisfy the classical
property of definite positivity~\cite{Salencon2001}.
Comparing equation~\ref{eq:contraintesolide} with the second
egality~\ref{eq:cauchyelasdefopure} allows to compute $\vvvv{L}$.
\begin{equation}
  \label{eq:L}
   \vvvv{L} =  -   \frac{n^2}{2} \int_{V_0} 
               \left(\v{R} \otimes
                \v{F} \otimes  \frac{d g_0}{d\v{R}} \cdot \mathbb{A}
             +   g_0  \left( \v{R} \otimes
                 \frac{d  \v{F}}{d\v{R}}  \cdot \mathbb{A}
             +   \v{F} \underset{1 \leftrightarrow 2}{\otimes} \mathbb{A}
              \right) \right)\,dV
\end{equation}
with $ \v{F} \underset{1 \leftrightarrow 2}{\otimes} \mathbb{A} = F_j
\mathbb{A}_{ik\ell} \v{e}_i \otimes \v{e}_j \otimes \v{e}_k \otimes
\v{e}_{\ell}$.
The elastic tensor $\Elas$ can be easily computed by combinig
equations~\ref{eq:cauchyelasdefopure} and~\ref{eq:L}.

In the sequel, it is assumed that the suspension is isotropic in
the undeformed configuration.
Then the radial distribution function reads
$g_0(\v{R})=g_0(|\v{R}|)=g_0(R)$. 
Moreover, it is also assumed that the interparticle forces are
central, which writes $\psi(\v{R}) = \psi(R)$. 
The interparticle forces read
\begin{equation}
  \label{eq:defF}
    \v{F}  =  {\displaystyle - \frac{1}{R} \frac{d \psi}{d R}
    \v{R}}
\end{equation}
Putting equation~\ref{eq:defF}  into expression~\ref{stressmacro} yields the
value of the stress tensor in the undeformed configuration 
\begin{equation}
\label{eq:defPi}
< \vv{\pi}^0>=\frac{2\pi}{3}n^2
\int_{R=2a}^{\infty}R^{3}\frac{d\psi(R)}{dR}g_0(R)dR \,\tun
\end{equation}
In equation~\ref{eq:defPi}, it was assumed that $R$ is unbounded
whereas one would expect that the r.e.v. is of finite extent.
Insofar as the r.e.v. must be large enough to be of typical
composition and that its overall properties do not depend ont its
size, the macroscopic behavior of the suspension can be defined only
if the decay of the interparticle forces for large  $R$ is strong enough that
the contribution of long range forces is negligible.
When these conditions are fulfilled, it is possible to simplify the
computation of quantities defined as an average over the r.e.v. by
assuming that the r.e.v. is unbounded.
The same truncature process is used in the sequel of the paper.
As the stress tensor in the reference configuration is isotropic,
the internal force is characterized by a pressure,  equal (up
to the fluctuation term $n k_B T$) to the osmotic pressure of a
colloidal system classically defined in the framework of statistical
mechanics~\cite{Russel-Saville-Schowalter:1995}.
Thanks to the fact that the initial configuration is isotropic and
the interparticle forces are central, we obtain
\begin{equation}
\label{eq:resultats}
{\displaystyle   \frac{d g_0}{d \v{R}}  }  =  {\displaystyle
  \frac{1}{R} \frac{d g_0}{d R} \v{R}} 
\end{equation}
The tensor field $\mathbb{A}$ depends upon the morphological
properties of the suspension in the undeformed configuration.
As it is not possible to compute $\mathbb{A}$ from a practical point
of view, we purpose to compute the overall properties of the suspension
using the classical choice $\forall \v{R}$, $\mathbb{A} (\v{R}) : \epsl =
\epsl \cdot \v{R}$ leading to the popular ``mean field theory''.
Using this particular localization
field  allows to obtain only estimates of the overall properties of
the suspension because this  choice
defines the solution of the problem under consideration only in
particular situation (uniform radial distribution function, periodic
lattice, \ldots)

Combining this estimate with the relations~\ref{eq:contraintesolide},
\ref{eq:resultats} and the incompressibility condition $\epsl : \tun
= 0$ yields the following behavior law of the suspension
\begin{equation}
<\vv{\tau}>  =  2G^{s} \epsl,
\hspace{0.5cm} \text{ with  } 
\hspace{0.5cm}
G^{s}  = \frac{3\phi^{2}}{40\pi
  a^{6}}\int_{R=2a}^{\infty}\frac{d}{dR}\left(R^{4}
\frac{d\psi}{dR}(X)g_0(R)\right)dR 
\label{eq:Gsolide}
\end{equation}
where $\phi=n4\pi a^{3}/3$ denotes the volumic fraction and $\vv{\tau}
= \cauchy - (\tun : \cauchy)/3 \tun$ the deviatoric part of the Cauchy
stress tensor.
It can be shown that the "solid modulus"~\ref{eq:Gsolide} is no more
than the classical Voigt estimate one can obtain using the uniform
strain field as a trial field in a variational approch to the problem
under consideration. Then, equation~\ref{eq:Gsolide} defines an upper bound
of the real solid shear modulus of the suspension.

It is reminded that classically, the actual configuration is taken as
the reference configuration to compute the elastic moduli
tensor~\cite{Wagner:1993}.
This result can be recovered by performing exactly the same
computations than above on the undeformed  configuration.
This approach yields the ``liquid'' elastic shear estimate
 \begin{equation}
<\vv{\tau}>  =  2G^{\ell} \epsl, 
\hspace{0.5cm} \text{ with  } 
\hspace{0.5cm}
\label{Gliquide} 
G^{\ell}  =  \frac{3\phi^{2}}{40\pi
  a^{6}}\int_{R=2a}^{\infty}\frac{d}{dR}
\left(R^{4}\frac{d\psi}{dR}(R)\right)g_0(R)dR
\end{equation}   
The difference between the two estimates comes from the fact that the
radial distribution function is derived with respect to $R$ in
equation~\ref{eq:Gsolide} and not in equation~\ref{Gliquide}.
It is possible to obtain the ``liquid'' estimate from the ``solid''
one by assuming the radial distribution function conservation in the
course of deformation.
Then, it is shown from equation~\ref{eq:contraintesolide} that, up
to the first order of~$\epsl$, the Cauchy stress tensor reads 
\begin{equation}
<\cauchy>=<
 \vv{\pi}^0> -\frac{n^{2}}{2}\int_{V_{0}} \frac{d\left[\v{R}
  \otimes \v{F} \right]}{d \epsl} : \epsl \,  g_{0}(\v{R}) dV(\v{R})
\label{eq:contrainteliquid}
\end{equation}
(it is always assumed that the material is incompressible).
Considering one more time that the suspension is isotropic in the
reference configuration, one readily obtains from
equation~\ref{eq:contrainteliquid} the estimate~\ref{Gliquide}
for the elastic shear modulus of the suspension in the framework of a
mean field theory.

\section{Conclusions}
\label{conclu}

We have obtained a new expression for the elastic shear modulus of a
colloidal suspension modeled as a solid. 
This expression allows to estimate the elastic modulus
of a yield stress suspension submitted to a load smaller than the yield
stress.
This result was obtained in the framework of an homogenization
approach to the behavior of a colloidal suspension considered as a
discrete solid medium.
Even if this approach relies on assumptions rather different from those
classically performed to obtain estimates for the overall properties
of suspensions in the framework of statistical mechanics, it is worth
noting that classical results can also be recovered.
Thus, we have shown that our estimate coincides with
the classical high-frequency modulus estimate when the actual configuration is
taken as the reference.
From our point of view, this result was recovered by modeling the
suspension as a ``liquid'', ie a suspension without a yield stress.
In this situation, it is not possible to define an ``undeformed''
configuration and the actual configuration is taken as the reference.
To our opinion, this similarity is a strong indication that both
approaches are consistent one to the other.

Furthermore, it has been recalled that the tangent modulus
is not equal to the elastic modulus when the stress applied to the material in
the reference configuration and the elastic modulus are of the same
order of magnitude.

LMSGC de l'Institut Navier (Universit\'e Paris Est) is UMR 113 of the
CNRS associated with the ENPC and the LCPC. 

Support from the Agence Nationale de la Recherche (ANR) is acknowledged (grant ANR-05-JCJC-0214).
% etc, etc

% \appendix

% \section{}
% \label{}

% The Acknowledgements are also a un-numbered section
%\section*{Acknowledgements}
% Acknowledgements text here


\begin{thebibliography}{13}
% please try to use the bibitem system -
% the references should be in the order they appear in the text and
%   there not not be any citations in the abstract or resume!

%\bibitem{label}



\bibitem{Zwanzig-Mountain:1965}
Zwanzig~R. and Mountain~R.~D.
\newblock {High-Frequency Elastic Moduli of Simple Fluids}
\newblock { J. Chem. Phys.} 43 (1965) 4464-4471.

\bibitem{Wagner:1993}
Wagner~N.~J.
\newblock {The High-Frequency Shear Modulus of Colloidal Suspensions
  and the Effects of Hydrodynamic Interactions}
\newblock { J. Colloid Interface Sci.} 161 (1993) 169-181.

\bibitem{Lionberger-Russel:1994}
Lionberger~R.~A.and Russel~W.~B.
\newblock {High frequency modulus of hard spher colloids}
\newblock { J. Rheology} 38 (1994) 1885-1908.

\bibitem{Brady:1993}
Brady~J.
\newblock {The rheological behavior of concentrated colloidal dispersions}
\newblock { J. Chem. Phys. 99 (1993) 567-581.}

\bibitem{Bergenholtz-Willenbacher-Wagner-Morrison-Ende-Mellema:1998}
Bergenholtz~J., Willenbacher~N., Wagner~N.~J., Morrison~van~den~Ende,
Mellema~J.
\newblock {Colloidal charge determination in concentrated liquid
  dispersions using torsional resonance oscillation}
\newblock { Journal of Colloid Interface Science 202 (1998) 430-440.}

\bibitem{Batchelor:1970}
Batchelor~G.
\newblock {The stress system in a suspension of force-free particles}
\newblock {J. Fluid Mech. 41 (1970) 545-570.}

\bibitem{Batchelor:1977}
Batchelor~G.
\newblock {The effect of brownian motion on the bulk stress in a suspension of
spherical particles}
\newblock {J. Fluid Mech. 83 (1977) 97-117.}

\bibitem{Russel-Saville-Schowalter:1995}
Russel~W.~B., Saville~D.~A. and Schowalter~W.~R.
\newblock {Colloidal Dispersions}
\newblock {Cambridge University Press, 1995.}

\bibitem{Chateau02}
Chateau~X., Moucheront~P. and Pitois~O.
\newblock {Micromechanics of unsaturated granular media}
\newblock {ASCE Journal of Engineering Mechanics 128(8) (2002) 856-863.}

\bibitem{Zaoui02}
Zaoui~A.
\newblock {Continuum Micromechanics: Survey}
\newblock {ASCE Journal of Engineering Mechanics 128(8) (2002) 808-816.}

\bibitem{Hansen-McDonald:1986}
J. P. Hansen~J.~P. and McDonald~Ian~R.
\newblock {Theory of simple liquids}
\newblock {Academic Press, London 1986.}

\bibitem{Germain73}
Germain~P.
\newblock {Cours de m\'ecanique des milieux continus},
\newblock { Masson, Paris, 1973.}

\bibitem{Salencon2001}
Salen\c{c}on~J.
\newblock {Handbook of continuum mechanics. General concepts,
  Thermoelasticity},
\newblock { Springer, Berlin, 2001.}

\end{thebibliography}
\end{document}
\begin{equation}
  \label{eq:L}
   \vvvv{L} =  -   \frac{n^2}{2} \int_{V_0} 
               \left(\v{R} \otimes
                \v{F} \otimes  \frac{d g_0}{d\v{R}} \cdot \mathbb{A}
             +   g_0  \left( \v{R} \otimes
                 \frac{d  \v{F}}{d\v{R}}  \cdot \mathbb{A}
             +   \v{F} \underset{1 \leftrightarrow 2}{\otimes} \mathbb{A}
              \right) \right)\,dV
\end{equation}
with $ \v{F} \underset{1 \leftrightarrow 2}{\otimes} \mathbb{A} = F_j
\mathbb{A}_{ik\ell} \v{e}_i \otimes \v{e}_j \otimes \v{e}_k \otimes
\v{e}_{\ell}$.
The elastic tensor $\Elas$ can be easily computed by combinig
equations~\ref{eq:cauchyelasdefopure} and~\ref{eq:L}.